\documentclass[a4paper,11pt]{article}
\usepackage[a4paper,includemp=false,body={16cm,24.7cm},vcentering]{geometry}
\usepackage[latin1]{inputenc}
\usepackage[T1]{fontenc}
\usepackage{mathptmx} 
\usepackage[bf,center]{caption}
\usepackage{graphicx} 

\usepackage{hyperref,amsmath,amssymb,amsthm}

\usepackage{multirow}

\makeatletter
\renewcommand{\section}{\@startsection{section}{1}{0mm}{30pt}{12pt}{\normalfont\normalsize\bfseries}}

\renewcommand{\subsection}{\@startsection{subsection}{2}{0mm}{18pt}{12pt}{\normalfont\normalsize\itshape}}
\makeatother
\newcommand{\Title}[1]{\begin{center}{\bfseries\fontsize{12pt}{12pt}\selectfont#1}\end{center}}
\newcommand{\Author}[2]{\begin{center}{\fontsize{12pt}{12pt}\selectfont#1}\\{\it #2~}\end{center}}
\newcommand{\Introduction}{\section*{Introduction}}
\newcommand{\Conclusion}{\section*{Conclusion}}

\begin{document}

\Title{Detection of inner Solar System Trojan Asteroids by Gaia}
  
\Author{M. Todd$^1$, P. Tanga$^2$, D.M. Coward$^3$, M.G. Zadnik$^1$}{1. Department of Imaging and Applied Physics, Bldg 301, Curtin University, Kent St, Bentley, WA 6102, Australia \\
2. Laboratoire Lagrange, UMR7293, Université de Nice Sophia-Antipolis, CNRS, Observatoire de la Côte d'Azur (France)\\
3. School of Physics, M013, The University of Western Australia, 35 Stirling Hwy, Crawley, WA 6009, Australia}

\subsubsection*{Abstract}

The Gaia satellite, planned for launch by the European Space Agency (ESA) in 2013, is the next generation astrometry mission following Hipparcos. While mapping the whole sky, the Gaia space mission is expected to discover thousands of Solar System Objects. These will include Near-Earth Asteroids and objects at Solar elongations as low as 45 degrees, which are difficult to observe with ground-based telescopes. We present the results of simulations for the detection of Trojan asteroids in the orbits of Earth and Mars by Gaia.
 
\Introduction

\noindent

Trojan asteroids share the orbit of a planet and librate about the L4 and L5 Lagrangian points in that planet's orbit. 
Earlier modelling and simulations for Earth Trojans \cite{mor02} predict the existence of $\sim 17$ bodies larger than 100 m, and for Mars Trojans \cite{tab99,tab00a,tab00b} $\sim 50$ bodies larger than 1 km are predicted.
The first discovery of a Trojan asteroid in Earth's orbit (2010 TK7) was announced in 2011 \cite{Connors11}, and there are only three known Mars Trojans. 
Based on those simulations it is possible that more Trojan asteroids exist in the inner Solar System, however detection of such objects by ground-based telescopes are subject to a number of restrictions. 
This paper describes the results of modelling and simulations for the detection of Trojan asteroids in the orbits of Earth and Mars by the Gaia space mission, which does not share those limitations of ground-based telescopes.

\section{Detection by Gaia}

\noindent

By definition, Trojan asteroids are found in the L4 and L5 Lagrangian regions in a planet's orbit. 
The probable orbits of Earth and Mars Trojans \cite{2012MNRAS.420L..28T,2012MNRAS.424..372T} show that peak detection longitudes are consistent with classical Lagrangian points, but that bodies are unlikely to be co-planar as the stable orbits have significant inclinations. The sky area in which these bodies may be located is quite large but this is not a significant problem in the context of Gaia's mission to survey the whole sky.

The apparent magnitude for an Earth Trojan ranges from $V=17.9$ to $V=19.5$ \cite{2012MNRAS.420L..28T} and for a Mars Trojan is between $V=16.2$ to $V=20.7$ \cite{2012MNRAS.424..372T}, assuming 1~km diameter and an albedo of 0.20, and neglecting atmospheric extinction. This variation is effected by phase angle and distance, with distance being the dominant influence. 
Since Earth Trojans are co-orbital then these would be brightest when nearest to Earth in their orbits. 
By comparison the Mars Trojans have greatest brightness at opposition, which is a region outside Gaia's scanning area. 
The brightness of the Mars Trojans, when within Gaia's scanning area, is $V>\sim18$.

\begin{figure}
\centering
\includegraphics[width=0.55\textwidth]{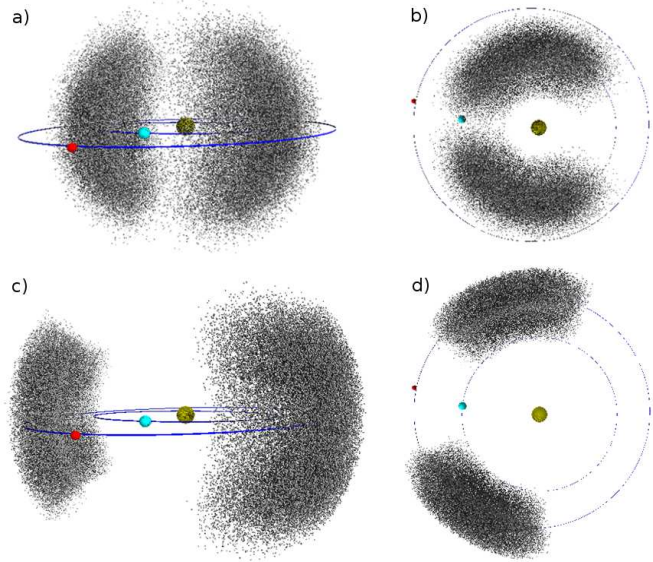}
\caption{Positions of simulated bodies at JD 2456000.5. a) Distribution of simulated Earth Trojans; b) Projection on the ecliptic plane of the positions of the simulated Earth Trojans; c) Distribution of Mars Trojans; and d) Projection on the ecliptic plane of the positions of the simulated Mars Trojans.} 
\label{fig1}
\end{figure}

The limits from the models were used to construct orbits for $20~000$ objects each for Earth and Mars Trojans, for simulation of possible detection by Gaia. The instantaneous positions of these are shown in Figure \ref{fig1}. 

For the simulated Earth Trojans, 969 orbits never crossed Gaia's field of view. A further 146 had brightnesses $V>20$, placing them below the detection threshold of $V=20$. The result of the simulation is that $\sim94$ per cent of objects were detected. 

For the simulated Mars Trojans, 142 orbits never crossed Gaia's field of view. In contrast to the detection results for the Earth Trojans, only 2096 were detected with brightness $V<20$. However $420$ of these were detected in only one telescope, which would prevent any orbit calculation and subsequent recovery. The result of the simulation is that $\sim8$ per cent of the objects were detected where possible follow-up study could be performed.

\begin{figure}
\centering
\includegraphics[width=\textwidth]{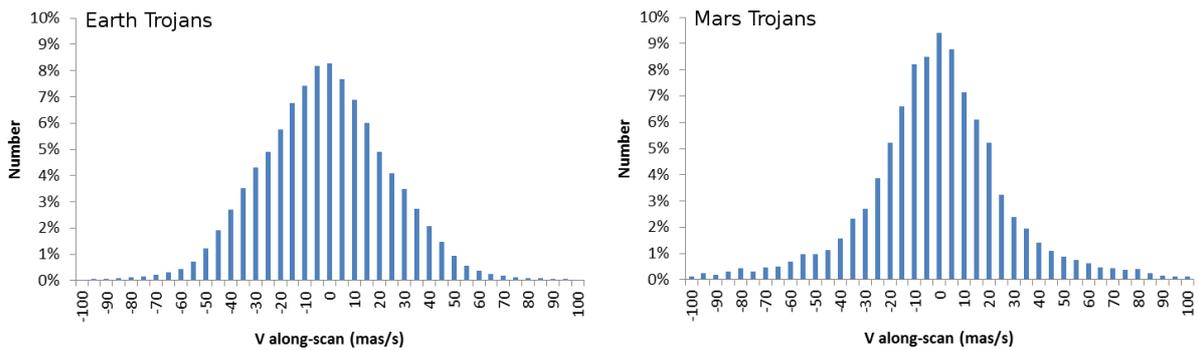}
\caption{Statistical distribution of the along-scan velocity during observation for the simulated objects.} 
\label{fig2}
\end{figure}

In both cases the along-scan velocity is significant. The along-scan velocity for Main Belt asteroids is typically less than $10$ mas/s \cite{Mignard11} whereas for both Earth Trojans and Mars Trojans along-scan velocities much greater than $10$ mas/s is quite common (Figure \ref{fig2}). The length of the window defined for sources $V>16$ is six pixels ($354$ mas). For an along-scan drift $>\sim3.5$ mas/s a source will travel $>175$ mas from the centre of the defined window in the along-scan direction in the $\sim50$ seconds it takes to travel the length of the CCD array, with the result that the source drifts outside its window. 

\begin{figure}
\centering
\includegraphics[width=\textwidth]{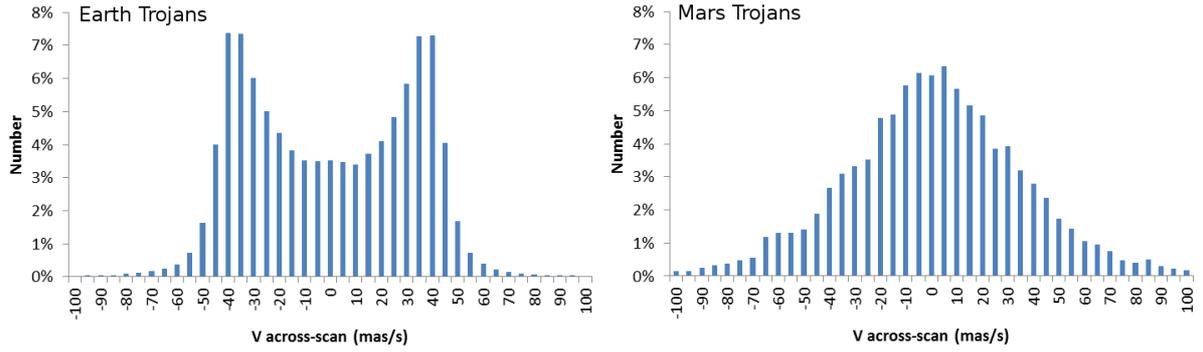}
\caption{Statistical distribution of the across-scan velocity during observation for the simulated objects.} 
\label{fig3}
\end{figure}

In both cases the across-scan velocity is also significant. The across-scan velocity for Main Belt asteroids is typically less than $15$ mas/s, whereas the across-scan velocities for Earth and Mars Trojans are much greater (Figure \ref{fig3}). The bimodal distribution of the across-scan velocities for Earth Trojans in Figure \ref{fig3} is a product of the geometry. The width of the window defined on the CCD is 12 pixels ($2124$ mas), and so an across-scan drift $>\sim21$ mas/s results in the source drifting outside the window. In addition, for across-scan velocity $>195$ mas/s, from a mean starting point at the centre of the CCD, the source will only be observed in one telescope.

Due to their orbits, the simulated Earth Trojans are observed more frequently than Main Belt whereas the simulated Mars Trojans are observed less frequently. This over-representation of the Earth Trojans in the detection statistics is expected; these are co-orbital and are observed at least once during each scanning cycle. However it is unlikely that there are any Earth Trojans of this size. The known Earth Trojan (2010~TK7) is estimated to have a diameter of $\sim300$~m \cite{Connors11}, with an apparent magnitude between $V=20.9$ to $V=22.7$, and a mean sky motion calculated to be between $25$~mas/s to $100$~mas/s (see Table \ref{tabasteroids}), making it unlikely to be detected by Gaia. By comparison the three Mars Trojans are much brighter and have a smaller apparent motion, making it probable that they will be detected by Gaia.

\begin{table}[tbh!]
\begin{center}
\caption{ \small Inner Solar System Trojan Asteroid magnitudes and mean sky motions.}
\label{tabasteroids}
\begin{tabular}{lcc}
\hline
Designation & Magnitude & Mean Sky Motion \\
            &           & (mas/s)   \\
\hline
2010~TK7    & $20.9<V<22.7$ & 25 - 100  \\
5261 Eureka & $17.1 < V < 19.2$ & 4.5 - 23.5   \\
1998 VF31   & $17.3 < V < 20.1$ & 6.5 - 35.5   \\
1999 UJ7    & $17.4 < V < 19.6$ & 4.5 - 23.0  \\
\hline
\end{tabular}
\end{center}
\end{table}

\Conclusion

The regions where Earth and Mars Trojan asteroids may be found occupy a very large sky area, however Gaia will survey these regions multiple times during its mission. Because of the co-orbital nature of Earth Trojans, that region will be over-observed by comparison with any other field since Gaia will observe that region each scan cycle.
In contrast with this, the Mars Trojan region will be observed much less frequently because of the different geometry.
For both cases the high along-scan and across-scan velocities may present a problem due to loss of signal as the source will tend to drift out of the window defined on the CCD.

It is unlikely that any Earth Trojans exist which are larger than 2010~TK7. The consequence is that any other Earth Trojans will be too small and hence too faint for Gaia to detect. In the case of the Mars Trojans there is some uncertainty but it is expected that Gaia would detect any additional bodies which are of sizes comparable to those already known. An analysis for the detection of the known Earth and Mars Trojans by Gaia will be forthcoming.

\section*{Acknowledgements} 
MT acknowledges support from the Astronomical Society of Australia, Australian Institute of Physics, and the sponsoring organisations of the Gaia-FUN-SSO2 workshop. MT thanks the SOC/LOC of the Gaia-FUN-SSO2 workshop for providing a fertile environment for discussing Gaia science. 
DMC is supported by an Australian Research Council Future Fellowship.

\end{document}